\documentclass[%
aps,%
showpacs,%
floatfix,%
superscriptaddress,%
nofootinbib,%
preprintnumbers,%
showkeys
]{revtex4}

\usepackage[dvips]{graphicx}
\usepackage{amsmath,amssymb}
\usepackage{color}

\preprint{STUPP-09-202, \\ }



\begin{document}

\title{Model building by coset space dimensional reduction in 
eight-dimensions}
\date{\today}

\author{Toshifumi Jittoh}
\email{jittoh@krishna.th.phy.saitama-u.ac.jp}
\affiliation{Department of Physics, Saitama University, 
        Shimo-Okubo, Sakura-ku, Saitama, 338-8570, Japan}
\author{Masafumi Koike}
\email{koike@krishna.th.phy.saitama-u.ac.jp}
\affiliation{Department of Physics, Saitama University, 
        Shimo-Okubo, Sakura-ku, Saitama, 338-8570, Japan}
\author{Takaaki Nomura}
\email{nomura@krishna.th.phy.saitama-u.ac.jp}
\affiliation{Department of Physics, Saitama University, 
        Shimo-Okubo, Sakura-ku, Saitama, 338-8570, Japan}
\author{Joe Sato}
\email{joe@phy.saitama-u.ac.jp}
\affiliation{Department of Physics, Saitama University, 
        Shimo-Okubo, Sakura-ku, Saitama, 338-8570, Japan}
\author{Yutsuki Toyama}
\email{toyama@krishna.th.phy.saitama-u.ac.jp}
\affiliation{Department of Physics, Saitama University, 
        Shimo-Okubo, Sakura-ku, Saitama, 338-8570, Japan}

\begin{abstract}
We investigate gauge-Higgs unification models in eight-dimensional
spacetime where extra-dimensional space
has the structure of a four-dimensional compact coset space.
The combinations of the coset space and the gauge group in the
eight-dimensional spacetime of such models are listed.
After the dimensional reduction of the coset space, we identified
$\mathrm{SO}(10)$, $\mathrm{SO}(10) \times \mathrm{U}(1)$ and
$\mathrm{SO}(10) \times \mathrm{U}(1) \times \mathrm{U}(1)$ as
the possible gauge groups in the four-dimensional theory that
can accomodate the Standard Model and thus is phenomenologically
promising.@
Representations for fermions and scalars for these
gauge groups are tabulated.

\end{abstract}

\maketitle

\section{Introduction}

The Standard Model has been eminently successful in describing the interactions
of the elementary particles.
A crucial role 
of this model is played by the Higgs scalar, which develops the
vacuum expectation value to
 give the masses to the elementary particles and
to trigger the breaking of the gauge symmetry from
\begin{math}
  \mathrm{SU}(3)_{\textrm{C}}
  \times \mathrm{SU}(2)_{\textrm{L}}
  \times \mathrm{U}(1)_{Y}
\end{math}
down to $\mathrm{SU}(3)_{\textrm{C}} \times \mathrm{U}(1)_{\textrm{EM}}$.
On the other hand, the most fundamental nature of the Higgs scalar such as its
mass is not predictable within the Standard Model.
Thus, 
the search for the nature of this particle is 
essential 
both for the confirmation of the Standard Model and for the search for new physics.

The gauge-Higgs unification is an attractive approach to account for the origin
of the Higgs scalars \cite{%
  Manton:1979kb,%
  Fairlie:1979at,%
  Fairlie:1979zy%
}(for recent approaches, 
 see Refs.~\cite{%
  Hall:2001zb,%
  Burdman:2002se,%
  Gogoladze:2003ci,%
  Scrucca:2003ut,%
  Haba:2004qf,%
  Biggio:2004kr,%
  Maru:2004,%
  Haba:2005kc,%
  Hosotani:2006qp,%
  Sakamoto:2006wf,%
  Maru:2006,%
  Hosotani:2007qw,%
  Sakamura:2007qz,%
  Medina:2007hz,%
  maru2007,%
  Adachi:2007tj,%
  Gogoladze:2007ey%
}).
This approach counts the Higgs scalars as components of the gauge bosons in the
spacetime with the dimension higher than four, and attributes their properties
to the physical setups such as the gauge symmetry and the compactification scale
of the extra-dimensional space.
We consider this idea in the scheme of the coset space dimensional reduction, in
which the extra-dimensional space is assumed to be a coset space of compact Lie
groups, and the gauge transformation is identified as the translation within
this space \cite{
Manton:1979kb,
Forgacs:1979zs,
Kapetanakis:1992hf,
Chatzistavrakidis:2007by,
Zoupanos08,
Jittoh:2008bs,
Jittoh:2008jc}.
This identification determines both the gauge symmetry and the particle contents
of the four-dimensional theory.

A phenomenologically promising gauge theory in a $D$-dimensional spacetime,
where $D > 4$, should reproduce the Standard Model after the dimensional
reduction.
Theories in six- and ten-dimensional spacetime has attracted much attention so
far in this regard.
The chiral structure of the matter content as in the Standard Model is easy to
introduce in these cases, more generally when $D = 4n + 2$
\cite{Chapline:1982wy,Chiral-con:C14}.
Fermions belonging to a vectorlike representation in ($4n + 2$)-dimensional
gauge theory can end up in a chiral fermion after dimensional reduction by
simultaneously applying the Weyl and the Majorana conditions, which are
compatible in this dimensionality.
This advantage increases the 
chance for the higher-dimensional model to be
a 
promising candidate.
No theories have been found quite promising, 
however,
for 
the 6, 10, 
and 14 dimensional 
spacetimes \cite{
  Kapetanakis:1992hf,%
  10dim-Model:F2,%
  10dim-Model:D3,%
  10dim-Model:K4,%
  10dim-Model:N5,%
  10dim-Model:K6,%
  10dim-Model:K12,%
  10dim-Model:D14,%
  10dim-Model:B,%
Jittoh:2008jc}.
%
%

We examine 
the theories in eight-dimensional spacetimes 
to search further for
promising theories.
The dimension of the extra-dimensional space $d = D - 4$ is four in this case,
and the small $d$ makes the problem tractable.
More importantly, the dimension $D = 8$ is below the critical dimension of the
string theories, which may thus supply the ultraviolet completions to models in
a spacetime of this dimensionality.
On the other hand, we need to confine ourselves to the complex representations
for the representations of the fermions, 
unlike the case of $D = 4n + 2$.

We search for the eight-dimensional gauge theory that leads to the
Standard Model, the GUTs, or their likes.
We exhaustively search for the possible candidates of coset space $S/R$ and the
gauge group $G$ of the eight-dimensional theory.
The representation for the gauge bosons is then automatically determined.
The representation of the fermions is searched up to 1000 dimensional ones,
while even larger representations are avoided lest it should generate numerous
unwanted fermions after the dimensional reduction.
We also tabulate the representation of the scalars and fermions under the gauge
group of the four-dimensional theory.

This paper is organized as follows.
In Section \ref{sec:Review}, we briefly recapitulate the scheme of the coset space dimensional reduction (CSDR) 
in eight dimensions. 
In Section \ref{sec:Search}, we search for the candidate of models in eight dimensions which lead to 
phenomenologically promising theories in four-dimensions after the dimensional reduction.
Section \ref{sec:Summary} is devoted to summary and discussions.

\section{CSDR scheme in eight dimensions}
\label{sec:Review}

In this section, we briefly recapitulate the scheme of the coset space
dimensional reduction in eight dimensions
~\cite{Kapetanakis:1992hf}.

We begin with a gauge theory defined on 
an eight-dimensional spacetime 
$M^{8}$
with a simple gauge group $G$.
Here $M^{8}$ is a direct product of a four-dimensional spacetime $M^4$ and 
a
compact coset space $S/R$, where $S$ is a compact Lie group and $R$ is a 
Lie
subgroup of $S$.
The dimension of the coset space $S/R$ is thus $4 \equiv 8 - 4$, implying
$\mathrm{dim} \, S - \dim \, R = 4$.
This structure of extra-dimensional space requires the group $R$ be 
embedded
into the group $\mathrm{SO}(4)$, which is a subgroup of the Lorentz group
$\mathrm{SO}(1, 7)$.
Let us denote the coordinates of $M^{8}$ by $X^{M} = (x^{\mu}, 
y^{\alpha})$,
where $x^{\mu}$ and $y^{\alpha}$ are coordinates of $M^{4}$ and $S/R$,
respectively.
The spacetime index $M$ runs over $\mu \in \{0, 1, 2, 3 \}$ and
$\alpha \in \{4, 5 , 6, 7 \}$.
In this theory, 
we introduce a gauge field $A_{M}(x, y) = (A_{\mu}(x, y),
A_{\alpha}(x, y))$, which belongs to the adjoint representation of the 
gauge
group $G$, and fermions $\psi(x, y)$, which lies in a representation $F$ of 
$G$.

The extra-dimensional space $S/R$ admits $S$ as an isometric transformation
group.
We impose on $A_{M}(X)$ and $\psi(X)$ the following symmetry under this
transformation in order to carry out the dimensional reduction
~\cite{Forgacs:1979zs,Symm-con:W13,Symm-con:J14,Symm-con:O15,Symm-con:Y16,Symm-con:Y16-2}.
Consider a coordinate transformation which acts trivially on $x$ and
gives rise to a $S$-transformation on $y$ as
$(x, y) \rightarrow (x, sy)$, 
where $s \in S$.
We require that the transformation of $A_{M}(X)$ and $\psi(X)$ under this
coordinate transformation 
be compensated by a gauge transformation.
This symmetry makes the eight-dimensional Lagrangian invariant under the
$S$-transformation and therefore independent of the coordinate $y$ of 
$S/R$.
The dimensional reduction is then carried out by integrating 
the eight-dimensional Lagrangian over the
coordinate $y$ to obtain the four-dimensional one.
The four-dimensional theory consists of the gauge fields $A_{\mu}$,
fermions $\psi$, and in addition the scalar fields originated from 
$A_{\alpha}$.
The gauge group reduces to a subgroup $H$ of the original gauge
group $G$.

The gauge symmetry and particle contents of the four-dimensional theory are
substantially constrained by the CSDR scheme.
We provide below the prescriptions to identify the four-dimensional gauge 
group
$H$ and its representations for the particle contents.
First, the gauge group of the four-dimensional theory $H$ is easily 
identified
as
\begin{equation}
  H = C_{G}(R),
\label{eq:H-is-centralizer}
\end{equation}
where $C_{G}(R)$ denotes the centralizer of $R$ in $G$~\cite{Forgacs:1979zs}.
Thus the four dimensional gauge group $H$ is determined by the embedding of 
$R$
into $G$.
These conditions imply 
\begin{eqnarray}
 G \supset H \times R,
\end{eqnarray}
up to U(1) factors.

Second, 
the representations of $H$ for the scalar fields are specified by 
the
following prescription.
Let us decompose the adjoint representation of $S$ according to the 
embedding $S
\supset R$ as,
\begin{align}
\label{dec-S-adj}
\mathrm{adj}~S
&= \mathrm{adj}~R + \sum_s {r}_s. 
\end{align}
We then decompose the adjoint representation of $G$ according 
to the
embeddings $G \supset H \times R$;
\begin{align}
\label{eq:dec_adjG_direct}
\mathrm{adj}~G &= (\mathrm{adj}~H, \mathbf{1}) + (\mathbf{1}, 
\mathrm{adj}~R) + \sum_{g} (h_g, {r}_g),
\end{align} 
where $r_{g}$s and $h_{g}$s denote representations of $R$ and $H$, 
respectively.
The representation of the scalar fields are $h_{g}$s whose
corresponding $r_{g}$s in the decomposition
Eq.~(\ref{eq:dec_adjG_direct}) are also 
contained in the set $\{ 
r_{s}\}$ in 
Eq.~(\ref{dec-S-adj}).

Third, 
the representation of $H$ for the fermion fields 
is
determined as follows~\cite{Manton:1981es}.
The $\mathrm{SO}(1,7)$ Weyl spinor $\mathbf{8}$ is decomposed under
its subgroup
\begin{math}
  (\mathrm{SU}(2)_{\mathrm{L}} \times \mathrm{SU}(2)_\mathrm{R}) (\simeq \mathrm{SO}(1, 3))
  \times (\mathrm{SU}(2)_1 \times \mathrm{SU}(2)_2) (\simeq \mathrm{SO}(4))
\end{math}
as
\begin{equation}
  \mathbf{8}
  = (\mathbf{2}_L, \mathbf{1}, \mathbf{2}_1,\mathbf{1})
  + (\mathbf{1}, \mathbf{2}_R, \mathbf{1},\mathbf{2}_2),
\end{equation} 
where $(\mathbf{2}_L, \mathbf{1})$ and $(\mathbf{1}, \mathbf{2}_R)$ 
representations
of $\mathrm{SU}(2)_\mathrm{L} \times \mathrm{SU}(2)_\mathrm{R}$ correspond to left- and 
right-handed
spinors, respectively.
The group $R$ is embedded into the Lorentz $(\mathrm{SO}(1,7))$ subgroup $\mathrm{SO}(4)$ 
in such a way that the vector representation $\mathbf{4}$ of 
$\mathrm{SO}(4)$ is
decomposed as 
$\mathbf{4} = \sum_{s} r_{s}$,
where $r_{s}$s
are the representations
obtained in the decomposition Eq.~(\ref{dec-S-adj}).
This embedding specifies a decomposition of the spinor representations
($\mathbf{2}_1, \mathbf{1}) ( (\mathbf{1}, {\mathbf{2}}_2) )$ of $\mathrm{SU}(2)_1\times\mathrm{SU}(2)_2 
\supset R$ as
\begin{equation}
  ( \mathbf{2}_1, \mathbf{1})  = \sum_{i} (\sigma_{1i})
 \quad \biggl(
    (\mathbf{1}, {\mathbf{2}_2}) = \sum_{i}
({\sigma}_{2i})
  \biggr).
\label{eq:sigmad-decomposition}
\end{equation}
We now decompose 
representation $F$ of the gauge group $G$
for the
fermions in eight-dimensional spacetime.
Decomposition of $F$ is
\begin{align}
\label{dec-F}
F &= \sum_f (h_f, {r}_f),
\end{align}
under $G \supset H \times R$.
The representations for the left-handed (right-handed) fermions are $h_{f}$s 
whose corresponding
$r_{f}$s are found in
\begin{math}
 \sigma_{1i}
  ( {\sigma}_{2i} )
\end{math}
obtained in Eq.~(\ref{eq:sigmad-decomposition}).

A phenomenologically acceptable model needs chiral fermions in 

four dimensions as the SM does.
%
%
The $\mathrm{SO}(1,7)$ spinor is not self-dual and 
its charge conjugate state is in a different representation
from itself. Thus the Majorana condition cannot be used to obtain a chiral structure from a vectorlike representation of $G$.
Therefore, 
we need to introduce complex representation for eight-dimensional fermions. 
Thus eight-dimensional model possesses
a completely different feature from $4n+2$-dimensional models.
We must 
work on complex representation for eight-dimensional
fermions.

Finally coset space $S/R$ of our interest 
should satisfy $\mathrm{rank} \, S =
\mathrm{rank} \, R$ to generate chiral fermions in four 
dimensions~\cite{Bott}.
We list all of four-dimensional coset spaces $S/R$ satisfying the condition 
and decompositions of SO(4) spinor and vector representation in Table~\ref{so4_vector_and_spinor}. 
%
%

%
\begin{table}
  \caption{%
    A complete list of four-dimensional coset spaces $S/R$ with 
    $\mathrm{rank} S = \mathrm{rank} R$.
    We also list the
    decompositions of the vector representation $\mathbf{4}$
    and the spinor representation $(\mathbf{2}_1,\mathbf{1}) $
    + $(\mathbf{1}, \mathbf{2}_2)$ of 
    $\mathrm{SO}(4)\simeq \mathrm{SU}(2)_1\times \mathrm{SU}(2)_2$
    under the $R$s.
    The representations of $r_{s}$ in Eq.~(\ref{dec-S-adj}) and 
    $\sigma_{1i}$  and $\sigma_{2i}$ in Eq.~(\ref{eq:sigmad-decomposition}) 
    are listed in the columns of ``Branches of $\mathbf{4}$'' and 
    ``Branches of $\mathbf{2}$'', respectively. 
    %
  }
  \label{so4_vector_and_spinor}
\begin{scriptsize}
\begin{center}
  \renewcommand{\arraystretch}{1.2}
\begin{tabular}{llll}
  \hline
  $S/R$
  &
  & Branches of $\mathbf{4}$
  & Branches of $\mathbf{2}$
  \\ \hline \hline
  (i) 
  & $\mathrm{Sp}(4) / [\mathrm{SU}(2) \times \mathrm{SU}(2)]$
  & ($\mathbf{2}, \mathbf{2})$ 
  & $(\mathbf{2}, \mathbf{1})$ and $(\mathbf{1}, \mathbf{2})$ 
  \\
  \hline
   (ii) 
  & $\mathrm{SU}(3) / [\mathrm{SU}(2) \times \mathrm{U}(1)]$
  & $\mathbf{2}(\pm 1)$ 
  & $\mathbf{2}(0)$ and $\mathbf{1}(\pm 1)$ 
  \\
  \hline
  (iii)
  & $\left(\mathrm{SU}(2) / \mathrm{U}(1) \right)^2$
  & ($\pm\mathbf{1}, \pm\mathbf{1})$ 
  & $(\pm\mathbf{1},0)$ and $(0,\pm\mathbf{1})$ 
  \\ 
  \hline
\end{tabular}
\end{center}
\end{scriptsize}
\end{table}


\section{Search for candidates}
\label{sec:Search}

In this section we search 
for realistic models in the CSDR scheme in eight-dimensions. 
First we investigate four-dimensional gauge group $H$ and 
higher-dimensional gauge group $G$ for 
each coset space of (i), (ii)
and (iii) listed in Table.~\ref{so4_vector_and_spinor}. 
The four-dimensional gauge groups acceptable for $H$ are listed in Table.~\ref{candidates_of_H_G}. 
%
This Table is obtained from the following considerations. 
\begin{enumerate}
\item
The number of $\mathrm{U}(1)$s in $H$ 
must be more than that in $R$, 
since the $\mathrm{U}(1)$s in $R$ are also part of its centralizer, i.e. part of $H$.
Therefore, 
the number of $\mathrm{U}(1)$s in $H$ 
must be 
$R$ or 
more. 
We thus exclude $\mathrm{SU}(5)$, $\mathrm{SO}(10)$, 
and $\mathrm{E}_6$ for coset space of (ii) 
and $G_\mathrm{SM}$, $\mathrm{SU}(5)$, $\mathrm{SO}(10)$, $\mathrm{E}_6$, 
$\mathrm{SU}(5) \times \mathrm{U}(1)$, $\mathrm{SO}(10) \times \mathrm{U}(1)$, 
and $\mathrm{E}_6 \times \mathrm{U}(1)$ for 
coset space of (iii).  

\item 
We also exclude $G_\mathrm{SM}$ for the coset space 
of (ii) and $G_\mathrm{SM} \times \mathrm{U}(1)$ for the coset space of (iii).
The hypercharges of the SM should be reproduced by the U(1) charges in R, 
which means that all the hypercharges 
must appear in the decomposition of SO(4) spinor. 
%
The dimension of the SO(4) spinor representation is however two, 
and hence more than two different values of U(1) charges are not available. 
%
%
Consequently, 
these cases never reproduce the five hypercharges of the SM fermions.

\item
We allow at most one extra $\mathrm{U}(1)$ in four-dimensional gauge group.
This excludes 
$G_\mathrm{SM} \times \mathrm{U}(1)$, 
$\mathrm{SU}(5) \times \mathrm{U}(1)$, 
$\mathrm{SO}(10) \times \mathrm{U}(1)$, 
$\mathrm{E}_6 \times \mathrm{U}(1)$, 
$G_\mathrm{SM} \times \mathrm{U}(1) \times \mathrm{U}(1)$, 
$\mathrm{SU}(5) \times \mathrm{U}(1) \times \mathrm{U}(1)$, 
$\mathrm{SO}(10) \times \mathrm{U}(1) \times \mathrm{U}(1)$, 
and $\mathrm{E}_6 \times \mathrm{U}(1) \times \mathrm{U}(1)$ 
for coset space of (i), 
and $G_\mathrm{SM} \times \mathrm{U}(1) \times \mathrm{U}(1)$, 
$\mathrm{SU}(5) \times \mathrm{U}(1) \times \mathrm{U}(1)$, 
$\mathrm{SO}(10) \times \mathrm{U}(1) \times \mathrm{U}(1)$, 
and $\mathrm{E}_6 \times \mathrm{U}(1) \times \mathrm{U}(1)$ 
for coset space of (ii). 

\end{enumerate}
The 
higher-dimensional gauge group $G$ should have 
the same rank 
as that of $H \times R$ up to U(1)s 
and 
possess complex representations to obtain chiral fermions. 
We also list the candidates of $G$ in Table~\ref{candidates_of_H_G}. 
\begin{table}[h]
\caption{
The candidates of $H$ and $G$. 
}
\label{candidates_of_H_G}
\begin{scriptsize}
\begin{center}
    \renewcommand{\arraystretch}{1.2}
    \begin{tabular}{lll} \hline
    $S/R$ & $H$ & $G$ \\ \hline \hline
    (i)
    & $\mathrm{SU}(3) \times \mathrm{SU}(2) \times \mathrm{U}(1)$ 
    & $\mathrm{SU}(7)$, $\mathrm{E}_6$ 
    \\ \cline{2-3}
    & $\mathrm{SU}(5)$ 
    & $\mathrm{SU}(7)$, $\mathrm{E}_6$ 
    \\ \cline{2-3}
    & $\mathrm{SO}(10)$ 
    & $\mathrm{SU}(8)$, $\mathrm{SO}(14)$ 
    \\ \cline{2-3}
    & $\mathrm{E}_6$ 
    & $\mathrm{SU}(9)$ 
    \\ \hline
    (ii) 
    & $\mathrm{SU}(3) \times \mathrm{SU}(2) \times \mathrm{U}(1) \times \mathrm{U}(1)$ 
    & $\mathrm{SU}(7)$, $\mathrm{E}_6$ 
    \\ \cline{2-3}
    & $\mathrm{SU}(5) \times \mathrm{U}(1)$ 
    & $\mathrm{SU}(7)$, $\mathrm{E}_6$ 
    \\ \cline{2-3}
    & $\mathrm{SO}(10) \times \mathrm{U}(1)$ 
    & $\mathrm{SU}(8)$, $\mathrm{SO}(14)$ 
    \\ \cline{2-3}
    & $\mathrm{E}_6 \times \mathrm{U}(1)$ 
    & $\mathrm{SU}(9)$ 
    \\ \hline
    (iii) 
    & $\mathrm{SU}(3) \times \mathrm{SU}(2) \times \mathrm{U}(1) \times \mathrm{U}(1) \times \mathrm{U}(1)$ 
    & $\mathrm{SU}(7)$, $\mathrm{E}_6$ 
    \\ \cline{2-3}
    & $\mathrm{SU}(5) \times \mathrm{U}(1) \times \mathrm{U}$ 
    & $\mathrm{SU}(7)$, $\mathrm{E}_6$ 
    \\ \cline{2-3}
    & $\mathrm{SO}(10) \times \mathrm{U}(1) \times \mathrm{U}$ 
    & $\mathrm{SU}(8)$, $\mathrm{SO}(14)$ 
    \\ \cline{2-3}
    & $\mathrm{E}_6 \times \mathrm{U}(1) \times \mathrm{U}$ 
    & $\mathrm{SU}(9)$ 
    \\ \hline
    \end{tabular}
\end{center}
\end{scriptsize}	   
\end{table}

We investigate representations of four-dimensional gauge group in the CSDR scheme. 
The representations for scalars in four-dimensional spacetime
are obtained by comparing Eq.~(\ref{dec-S-adj}) and Eq.~(\ref{eq:dec_adjG_direct}), 
while those for fermions  
are similarly obtained by comparing Eq.~(\ref{eq:sigmad-decomposition}) and Eq.~(\ref{dec-F}). 
Note again that $F$ must be complex representation in order to obtain chiral fermions in four dimensions.  
We limit the dimension of $F$ 
to 1000 to avoid numerous representations 
of the fermions under four-dimensional gauge group.

Exhaustive investigation of all combinations of $S/R$, $G$, $H$ and $F$ 
leaves six candidates of models which include at least one generation of 
known fermions; they are listed in Table~\ref{table_results}.  
\begin{table}[h]
\begin{scriptsize}
\renewcommand{\arraystretch}{1.2}
\begin{center}
\caption{
Four-dimensional scalar and fermion representations for each combination of $S/R$, $G$, $H$ and $F$. 
}
\label{table_results}
\begin{tabular}{l|l|l|l|l|l}\hline 
	$S/R$ & $H$ & $G$ & scalar & $F$ & fermions  \\ \hline \hline
	(i) 
	& $\mathrm{SO}(10)$ 
	& $\mathrm{SO}(14)$ & $\mathbf{10}$ 	& $\mathbf{64}$ 
	& $\{\mathbf{16}\}^2$ \\ \cline{5-6}
	& 					
	&					&					& $\mathbf{832}$ 	
	& $\{\mathbf{16}\}^2$, $\{\overline{\mathbf{144}}\}^2$ \\ \hline 
	(ii) 
	& $\mathrm{SO}(10) \times \mathrm{U}(1)$ 
	& $\mathrm{SO}(14)$ 	& $\mathbf{10}(1)$, $\mathbf{10}(-1)$ 	& $\mathbf{64}$ 	
	& $\mathbf{16}(0)$, $\mathbf{16}(1)$, $\mathbf{16}(-1)$ \\ \cline{5-6}
	&										
	&						& 										& $\mathbf{832}$ 	
	& $\{\mathbf{16}(0)\}^2$, $\mathbf{16}(1)$, $\mathbf{16}(-1)$, \\ 
	&										
	&						&										&
	& $\overline{\mathbf{144}}(0)$, $\overline{\mathbf{144}}(1)$, $\overline{\mathbf{144}}(-1)$ \\ \hline
	(iii) 
	& $\mathrm{SO}(10) \times \mathrm{U}(1) \times \mathrm{U}(1)$	
	& $\mathrm{SO}(14)$		& $\mathbf{10}(1,1)$, $\mathbf{10}(1,-1)$	& $\mathbf{64}$ 
	& $\mathbf{16}(1,0)$, $\mathbf{16}(-1,0),$ \\
	&
	& 						&$\mathbf{10}(-1,1)$, $\mathbf{10}(-1,-1)$ 	& 
	& $\mathbf{16}(0,1)$, $\mathbf{16}(0,-1)$  \\ \cline{5-6}
	&
	&						&											& $\mathbf{832}$ 
	& $\{\mathbf{16}(1,0)\}^2$, $\{\mathbf{16}(-1,0)\}^2,$ \\  
	&
	&						&											&
	& $\{\mathbf{16}(0,1)\}^2$, $\{\mathbf{16}(0,-1)\}^2,$ \\
	&
	&						&											&
	& $\overline{\mathbf{144}}(1,0)$, $\overline{\mathbf{144}}(-1,0),$ \\
	&
	&						&											&
	& $\overline{\mathbf{144}}(0,1)$, $\overline{\mathbf{144}}(0,-1)$ \\ \hline
\end{tabular}
\end{center}
\end{scriptsize}
\end{table}

For coset space of (i), we embed $R = \mathrm{SU}(2) \times \mathrm{SU}(2)$ into $G = \mathrm{SO}(14)$ according to the decomposition
\begin{equation}
\mathrm{SO}(14) \supset \mathrm{SU}(2) \times \mathrm{SU}(2) \times \mathrm{SO}(10). 
\label{decomposition_case1}
\end{equation} 
The decomposition of the adjoint representation of $\mathrm{SO}(14)$ according to the decomposition of
Eq.~(\ref{decomposition_case1}) is 
\begin{equation}
\mathbf{91}
=
(\mathbf{1},\mathbf{1},\mathbf{45})
+(\mathbf{3},\mathbf{1},\mathbf{1})
+(\mathbf{1},\mathbf{3},\mathbf{1})
+(\underline{\mathbf{2},\mathbf{2}},\mathbf{10}),
\end{equation}
and thus we obtain $\mathbf{10}$ as the scalar reperesentation in four dimensions.
Similarly, we decompose the complex representations $\mathbf{64}$ and $\mathbf{832}$ of $\mathrm{SO}(14)$ 
according to the decomposition of Eq.~(\ref{decomposition_case1}) as
\begin{align}
\mathbf{64}
=&
(\underline{\mathbf{2},\mathbf{1}},\mathbf{16})
+(\underline{\mathbf{1},\mathbf{2}},\overline{\mathbf{16}}), 
\\[10pt] 
\mathbf{832}
=&
(\underline{\mathbf{1},\mathbf{2}},\mathbf{144})
+(\underline{\mathbf{2},\mathbf{1}},\overline{\mathbf{144}})
+(\mathbf{2},\mathbf{3},\mathbf{16})
+(\mathbf{3},\mathbf{2},\overline{\mathbf{16}}) 
\nonumber \\
&+(\underline{\mathbf{1},\mathbf{2}},\overline{\mathbf{16}})
+(\underline{\mathbf{2},\mathbf{1}},\mathbf{16}),
\end{align}
and obtain 
$\{\mathbf{16}\}^2$ from $F=\mathbf{64}$ 
and $\{\mathbf{16}\}^2$ $+$ $\{\overline{\mathbf{144}}\}^2$ from $F=\mathbf{832}$  
as representations for the left-handed fermion in four dimensions.

For coset space of (ii), we embed $\mathrm{SU}(2) \times \mathrm{U}(1)$ into $\mathrm{SO}(14)$ according to the decomposition
\begin{align}
\mathrm{SO}(14) 
&\supset \mathrm{SU}(2) \times \mathrm{SU}(2) \times \mathrm{SO}(10) 
\nonumber \\
&\supset \mathrm{SU}(2) \times \mathrm{U}(1) \times \mathrm{SO}(10). 
\label{decomposition_case2}
\end{align} 
The decomposition of the adjoint representation of $\mathrm{SO}(14)$ according to the decomposition of 
Eq.~(\ref{decomposition_case2}) is 
\begin{align}
\mathbf{91} 
=& 
(\mathbf{1},\mathbf{1},\mathbf{45})
+(\mathbf{3},\mathbf{1},\mathbf{1})
+(\mathbf{1},\mathbf{3},\mathbf{1})
+(\mathbf{2},\mathbf{2},\mathbf{10}) 
\nonumber \\
=&
(\mathbf{1},\mathbf{45})(0)
+(\mathbf{3},\mathbf{1})(0)
+(\mathbf{1},\mathbf{1})(2)
+(\mathbf{1},\mathbf{1})(0)
+(\mathbf{1},\mathbf{1})(-2)
\nonumber \\
&+(\underline{\mathbf{2}},\mathbf{10})(\underline 1)
+(\underline{\mathbf{2}},\mathbf{10})(\underline{-1}), 
\end{align}
and thus we obtain $(\mathbf{10}(1))$ and $(\mathbf{10}(-1))$ as the scalar representations in four dimensions.  
Similarly, we decompose the complex representations $\mathbf{64}$ and $\mathbf{832}$ of $\mathrm{SO}(14)$ 
according to the decomposition of Eq.~\ref{decomposition_case2}) as  
\begin{align}
\mathbf{64} 
=& 
(\mathbf{2},\mathbf{1},\mathbf{16})
+(\mathbf{1},\mathbf{2},\overline{\mathbf{16}})
\nonumber \\
=& 
(\mathbf{2},\mathbf{16})(0)
+(\mathbf{1},\overline{\mathbf{16}})(1)
+(\mathbf{1},\overline{\mathbf{16}})(-1)
\\[10pt]
\mathbf{832} 
=& 
(\mathbf{1},\mathbf{2},\mathbf{144})
+(\mathbf{2},\mathbf{1},\overline{\mathbf{144}})
+(\mathbf{2},\mathbf{3},\mathbf{16})
+(\mathbf{3},\mathbf{2},\overline{\mathbf{16}})
\nonumber \\
&+(\mathbf{1},\mathbf{2},\overline{\mathbf{16}})
+(\mathbf{2},\mathbf{1},\mathbf{16}) 
\nonumber \\
=& 
(\underline{\mathbf{1}},\mathbf{144})(\underline{1})
+(\underline{\mathbf{1}},\mathbf{144})(\underline{-1})
+(\underline{\mathbf{2}},\overline{\mathbf{144}})(\underline{0})
+(\mathbf{2},\mathbf{16})(2)
\nonumber \\
&+(\underline{\mathbf{2}},\mathbf{16})(\underline{0})
+(\mathbf{2},\mathbf{16})(-2)
+(\mathbf{3},\overline{\mathbf{16}})(1)
+(\mathbf{3},\overline{\mathbf{16}})(-1)
\nonumber \\
&+(\underline{\mathbf{1}},\overline{\mathbf{16}})(\underline{1})
+(\underline{\mathbf{1}},\overline{\mathbf{16}})(\underline{-1})
+(\underline{\mathbf{2}},\mathbf{16})(\underline{0}),    
\end{align}
and obtain 
$\mathbf{16}(0)$, 
$\mathbf{16}(1)$ 
and $\mathbf{16}(-1)$ 
from $F=\mathbf{64}$ 
and 
$\{\mathbf{16}(0)\}^2$, 
$\mathbf{16}(1)$, 
$\mathbf{16}(-1)$, 
$\overline{\mathbf{144}}(0)$, 
$\overline{\mathbf{144}}(1)$ 
and $\overline{\mathbf{144}}(-1)$ 
from $F=\mathbf{832}$  
as representations for the left-handed fermion in four dimensions.

For coset space of (iii), we embed $\mathrm{U}(1)\times \mathrm{U}(1)$ into $\mathrm{SO}(14)$ according to the decomposition
\begin{align}
\mathrm{SO}(14) 
&\supset \mathrm{SU}(2) \times \mathrm{SU}(2) \times \mathrm{SO}(10)
\nonumber \\
&\supset \mathrm{SO}(10) \times \mathrm{U}(1) \times \mathrm{U}(1). 
\label{decomposition_case3}
\end{align} 
The decomposition of the adjoint representation of $\mathrm{SO}(14)$ according to the decomposition of 
Eq.~(\ref{decomposition_case3}) 
is 
\begin{align}
\mathbf{91} 
=& 
(\mathbf{1},\mathbf{1},\mathbf{45})
+(\mathbf{3},\mathbf{1},\mathbf{1})
+(\mathbf{1},\mathbf{3},\mathbf{1})
+(\mathbf{2},\mathbf{2},\mathbf{10})
\nonumber \\
=& 
\mathbf{45}(0,0)
+\mathbf{1}(2,0)
+\mathbf{1}(0,0)
+\mathbf{1}(-2,0)
+\mathbf{1}(0,2)
+\mathbf{1}(0,0)
\nonumber \\
&+\mathbf{1}(0,-2) 
+\mathbf{10}(1,1)
+\mathbf{10}(1,-1)
+\mathbf{10}(-1,1)
+\mathbf{10}(-1,-1), 
\end{align}
and thus we obtain $(\mathbf{10}(1,1))$, $(\mathbf{10}(1,-1))$, $(\mathbf{10}(-1,1))$ and $(\mathbf{10}(-1,-1))$ 
as the scalar representations in four dimensions.  
Similarly, we decompose the complex representations $\mathbf{64}$ and $\mathbf{832}$ of $\mathrm{SO}(14)$ 
according to the decomposition of Eq.~(\ref{decomposition_case3}) as 
\begin{align}
\mathbf{64} 
=& 
(\mathbf{2},\mathbf{1},\mathbf{16})
+(\mathbf{1},\mathbf{2},\overline{\mathbf{16}})
\nonumber \\
=& 
\mathbf{16}(1,0)
+\mathbf{16}(-1,0)
+\overline{\mathbf{16}}(0,1)
+\overline{\mathbf{16}}(0,-1)
\\[10pt]
\mathbf{832} 
=& 
(\mathbf{1},\mathbf{2},\mathbf{144})
+(\mathbf{2},\mathbf{1},\overline{\mathbf{144}})
+(\mathbf{2},\mathbf{3},\mathbf{16})
+(\mathbf{3},\mathbf{2},\overline{\mathbf{16}})
+(\mathbf{1},\mathbf{2},\overline{\mathbf{16}})
\nonumber \\
&+(\mathbf{2},\mathbf{1},\mathbf{16}) 
\nonumber \\
=& 
\mathbf{144}(0,1)
+\mathbf{144}(0,-1)
+\overline{\mathbf{144}}(1,0)
+\overline{\mathbf{144}}(-1,0) 
\nonumber \\
&+\mathbf{16}(1,2)
+\mathbf{16}(1,0)
+\mathbf{16}(1,-2)
+\mathbf{16}(-1,2) 
\nonumber \\
&+\mathbf{16}(-1,0)
+\mathbf{16}(-1,-2)
+\overline{\mathbf{1}6}(2,1)
+\overline{\mathbf{16}}(2,-1) 
\nonumber \\
&+\overline{\mathbf{16}}(0,1)
+\overline{\mathbf{16}}(0,-1)
+\overline{\mathbf{16}}(-2,1)
+\overline{\mathbf{1}6}(-2,-1) 
\nonumber \\
&+\overline{\mathbf{16}}(0,1)
+\overline{\mathbf{16}}(0,-1)
+\mathbf{16}(1,0)
+\mathbf{16}(-1,0), 
\end{align}
and obtain 
$\mathbf{16}(1,0)$, 
$\mathbf{16}(-1,0)$, 
$\mathbf{16}(0,1)$ 
and $\mathbf{16}(0,-1)$ 
from $F=\mathbf{64}$ 
and 
$\{\mathbf{16}(1,0)\}^2$, 
$\{\mathbf{16}(-1,0)\}^2$, 
$\{\mathbf{16}(0,1)\}^2$, 
$\{\mathbf{16}(0,-1)\}^2$, 
$\overline{\mathbf{144}}(1,0)$, 
$\overline{\mathbf{144}}(-1,0)$, 
$\overline{\mathbf{144}}(0,1)$ 
and $\overline{\mathbf{144}}(0,-1)$ 
from $F=\mathbf{832}$  
as representations for the left-handed fermion in four dimensions.

We can obtain 
one generation of the SM fermion from all of the candidates 
listed in Table~\ref{table_results} 
since the representations $\mathbf{16}$ and $\mathbf{144}$ of $\mathrm{SO}(10)$ include one generation of the SM fermion.  
The models with SU(3)/SU(2)$\times$U(1) are particularly intersting in our results.
%
We obtain 
the three 
generations of the SM fermions for this coset space with $F=\mathbf{64}$. 
%
We also obtain odd number generation in the combination of coset space of (ii) with $F=\mathbf{832}$. 
This is due to the fact 
that the SO(4) spinor is not self conjugate which forbid Majorana-Weyl condition 
and 
that $R$, 
which is SU(2)$\times$U(1), 
are embedded into SO(4) lopsidedly.

\section{Summary and discussions}
\label{sec:Summary}

We analyzed 
the gauge-Higgs unification models based on eight-dimensional gauge theories under the 
coset space dimensional reduction and exhaustively searched for models which lead to a phenomenologically promising
model after dimensional reduction.

We first made a complete list of the eight-dimensional models by determining the structure of the 
coset space $S/R$, the gauge group $G$, and the representations $F$ of $G$ for fermions.
We obtained a full list of the possible coset space $S/R$ in Table~{\ref{so4_vector_and_spinor}} 
by requiring dim$S/R$ =4 and rank$S$=rank$R$.
The gauge group $G$ is required to have complex representations and to lead to one of the 
following two-types of gauge groups after dimensional reduction: the GUT-like gauge groups such as E$_6$,
SO(10), SU(5) and these groups with one or two extra U(1)s, or the Standard-Model (SM)-like groups which are 
$G_{SM}=$SU(3)$\times$SU(2)$\times$U(1) and $G_{SM}$ with one or two extra U(1)s (see table.~\ref{candidates_of_H_G}).
The representation $F$ for fermions must be a complex representation of $G$ and is limited to be dimension  
less than 1000.

We then analyzed the particle contents of the four-dimensional theories that are 
obtained from each of the candidates ($S/R$,$G$,$F$). 
We found the phenomenologically promising models which induce $H$=SO(10)($\times$ one or two U(1)s) 
in four-dimensions, while other cases are found to be unsuccessful such as
$H$=SU(3)$\times$SU(2)$\times$U(1), $H$=SU(5), 
and $H$=E$_6$(with one or two extra U(1)s) 
in four-dimensions. 
%
We summarized these models in Table
~{\ref{table_results}}.
The SO(10) GUT-like models provide more than two fermions of $\mathbf{16}$ and $\mathbf{144}$ representations, 
along with a number of scalars of $\mathbf{10}$ representation.
A scalar of $\mathbf{10}$ can be interpreted as the electroweak Higgs particles.
More than two fermions of $\mathbf{16}$ and $\mathbf{144}$ representations would provide 
the generations of the fermions in the SM.
The most interesting model in this regard is the one with $S/R$ = SU(3)/SU(2)$\times$U(1), $G$=SO(14), 
$F$=$\mathbf{64}$ and $H$=SO(10)$\times$U(1). 
This model leads three fermions of $\mathbf{16}$, suggesting the existence of
three generations 
of the SM fermions.

An apparent challenge is 
to break the SO(10) gauge symmetry down to 
the SM ones.
This difficulty can be overcome by employing
the Hosotani mechanism, also known as the Wilson flux breaking mechanism 
\cite{Hosotani:1983xw,Hosotani:1983vn,W-breaking:E8}, or 
non-trivial boundary conditions of $S/R$ \cite{Nomura:2008sx}.
%
We leave further analysis for 
future study.
 
\acknowledgments

The work of T.~J. was financially supported by the Sasakawa Scientific Research 
Grant from the Japan Science Society. 
The work of T.~N. was supported in part by the Grant-in-Aid for the Ministry
of Education, Culture, Sports, Science, and Technology, Government of
Japan (No. 19010485). 
The work of J.~S. was supported in part by the Grant-in-Aid for the Ministry
of Education, Culture, Sports, Science, and Technology, Government of
Japan (No. 20025001, 20039001, and 20540251).


\end{document}